\documentclass[12pt,draft,aps,showpacs,nofootinbib,notitlepage,eqsecnum, floatfix,superscriptaddress]{revtex4}
\usepackage{amsmath}
\usepackage{amsfonts}
\usepackage{amssymb}
\usepackage{mathrsfs}
\usepackage{color}
\usepackage{graphicx}

\def\bc{\begin{center}}
\def\nno{\nonumber}
\def\ec{\end{center}}
\def\be{\begin{eqnarray}}
\def\ee{\end{eqnarray}}

\newcommand{\omits}[1]{}

\definecolor{dyellow}{rgb}{1.,0.8,.0}
\definecolor{myblue}{rgb}{.1,.1,.7}
\definecolor{dcyan}{rgb}{.0,.6,.6}
\definecolor{dmagenta}{rgb}{0.6,0.0,0.6}
\definecolor{brown}{rgb}{0.6,0.2,0.}
\definecolor{darkblue}{rgb}{.0,.0,0.5}
\definecolor{darkred}{rgb}{0.75,0.0,0.0}
\definecolor{orange}{rgb}{1.,.6,.0}
\definecolor{dorange}{rgb}{0.8,.4,.0}
\definecolor{darkgreen}{rgb}{0.0,0.6,0.0}
\definecolor{purple}{rgb}{.4,.0,.4}

\def\N{N \hspace{-0.7em}_{_{\sim}} ~}

\def\dl{\delta}
\def\eps{\epsilon}
\def\ka{\kappa}
\def\la{\lambda}

\def\si{\sigma}

\def\d#1#2{\frac{\displaystyle #1}{\displaystyle #2}}
\def\r{\partial}

\newcommand{\vect}[1]{\mbox{\boldmath $#1$}}

\newcommand\btd{\raise 2pt
\hbox{$\hat\bigtriangledown$}\hskip 1.5pt}
\newcommand\bt{\raise 2pt
\hbox{$\bigtriangledown$}\hskip 1.5pt}

\newcommand{\SR}{$SR$}
\newcommand{\GR}{$GR$}

\newcommand{\dS}{$d{S}$}

\newcommand{\AdS}{${A}d{S}$}

\newcommand{\PoR}{${P}o{R}$}
\newcommand{\PoRcl}{${P}o{R}_{c l}$}

\newcommand{\PDG}
{projective differential geometry}
\newcommand{\LFT}{$LFT$}
\newcommand{\R}{$\cal R$}

\newcommand{\Mink}{$Mink$}

\def\P{{\bf P}}
\def\K{{\bf K}}
\def\H{{H}}
\def\J{{\bf J}}
\def\N{{\bf N}}

\def\R{{\it R}}

\def\PRD{{\it Phys. Rev.}~{\bf D}}
\def\PRL{{\it Phys. Rev. Lett }}
\def\PLA{{\it Phys. Lett.}~{\bf A}}
\def\PLB{{\it Phys. Lett.}~{\bf B}}

\begin{document}

\title{\Large{Poincar\'e-De Sitter Flow and Cosmological Meaning}\footnote{Based partially on
HYG's invited talk at  Workshop on Lie Algebras in Physics, Hua L.K.
Key Lab. of Math. and Morningside Center of Math., CAS June 29-July
10, 2009; lectures at International Summer School on General
Relativity and Quantum Gravity, July 19-Aug. 1; and invited talk at
the Loops 2009, Aug. 2-7, 2009, Beijing Normal University, Beijing,
China.}}

\author{Han-Ying Guo}

\email{hyguo@itp.ac.cn}
\affiliation{ Institute of Theoretical Physics and Kavli Institute of
Theoretical Physics of China, Chinese Academy of Sciences,
   Beijing 100190, China}

\author{Hong-Tu Wu}
\email{htwu@amss.ac.cn}

\affiliation{Institute of Mathematics, Academy of Mathematics  and
System Sciences, Chinese Academy of Sciences,
   Beijing 100190, China}

   \author{YU Yue}
\email{yuyue@ihep.ac.cn} \affiliation{Theoretical Physics Center for
Science Facilities, Chinese Academy of Sciences,}
\affiliation{Institute of High Energy Physics,  Chinese Academy of
Sciences,
   Beijing 100049, China.}

\date{May, 2010}

\begin{abstract}
We introduce the Poincar\'e-de Sitter flow with real numbers
$\{r,s\}$ to parameterize the relativistic quadruple ${\frak
Q}_{PoR}=[{\cal P}, {\cal P}_2, {\cal D}_+,{\cal
D}_-]_{M/M_\pm/D_\pm}$ for the triple of  Poincar\'e/\dS/\AdS\ group
${\cal P}/{\cal D}_+/{\cal D}_-$ invariant special relativity. The dual
Poincar\'e group ${\cal P}_2$-invariant degenerated Einstein manifold
$M_\pm$ of $\Lambda_\pm=\pm3l^{-2}$ is for the space/time-like domain $\dot R_\pm$  of  the compact lightcone $\bar C_O$ associated to the common
space/time-like region $R_\pm$ of the lightcone $C_O$ at  common origin on
Minkowski/\dS/\AdS\ spacetime $M/D_+/D_-$. Based on the
principle of relativity with two universal constants $(c, l)$, there
are the law of inertia, coordinate time simultaneity and so on for the flow on a
Poincar\'e-\dS\ symmetric Einstein manifold of
$\Lambda_{s}=3{s}l^{-2}$. Further, there is Robertson-Walker-like
cosmos of the flow  for  the propertime simultaneity.
The \dS\ special relativity with double $[{\cal D}_+,{\cal
P}_2]_{D_+/M_+}$ can  provide a consistent kinematics for the
 cosmic scale  physics with an upper entropy bound $S_R=k_B\pi g^{-2}, g^2:=(\ell_P/R)^2
 \simeq 10^{-122}$, for $R\simeq (3/\Lambda)^{1/2}\sim 13.7 Gly$.
\end{abstract}


\maketitle


\tableofcontents


\section{Introduction}\label{sec:Intr}

Precise cosmology shows that our universe is accelerated expanding
 with a tiny positive cosmological constant $\Lambda_+$ and quite
possibly asymptotic to a Robertson-Walker-de Sitter
(\dS) space{time}.
This opens up the era of the cosmic scale physics with new
kinematics and dynamics characterized by the $\Lambda_+$. In Einstein's theory of relativity \cite{1905, 1922},  special relativity (\SR) is on the flat Minkowski (\Mink) spacetime,  general relativity   (\GR) is on the curved spacetime with the \Mink-spacetime as a tangent space and the cosmological constant $\Lambda_+$ is put in by hand. For the kinematics at the cosmic scale with $\Lambda_+$, Einstein's theory has to be re-examined from the very beginning: the principles. In particular, the principle of relativity (\PoR) in Einstein's \SR. \omits{ In bath cases, the Euclidean assumption for local ruler and clock has been taken so that the infinity cannot be reached.}

 As was well-known, the \PoR\ and the postulate on invariance of light velocity $c$ lead to Einstein's \SR\ on the
Poincar\'e invariant \Mink\ spacetime $M$. Inherited from Newton's
theory, the space and time of the \Mink-spacetime are Euclidean and the projective
infinity be excluded. The Euclidean assumption for  rigid ruler and ideal clock are reasonable
in conventional scales.  It is priori, however, for the cosmic scale kinematics. In order to avoid the Euclidean assumption and to allow
 the inertial motion of Newton's first
law may reach the (projective) infinity as a projective
straightline, the \PoR\ and the postulate should be extended and
weakened  to the \PoR\ with two universal constants $(c,l)$ (denoted
as \PoRcl). In fact, in addition to  $c$, the other universal
constant $l$ of length dimension should be introduced to
characterize the dimension of space coordinates and to link the
\PoR\ with the cosmological constant $\Lambda_+$.

If the inertial motion is allowed   may reach the infinity, the most
general transformations  among inertial frames can be found first
{as} the linear fractional transformations with common denominator
(\LFT s) \cite{Umov, Weyl, Fock, Hua, HuaC}, which may send finite
events to the infinity. And it can be classified afterwards what
kinds of kinematical symmetries of space and time may be included
for real physics with metric~\cite{SRT,SRT2}. This is quite
different from conventional understanding. Actually, these \LFT s
form the inertial motion group $IM(1,3)$ homomorphic to 4d real
projective group $PGL(5,R)$ with the inertial motion algebra
$\frak{im}(1,3)$ isomorphic to 4d real projective algebra
$\frak{pgl}(5,R)$. The non-orientation  of 4d real projective space
can be avoided without taking antipodal identification of the
inhomogeneous projective coordinates $x^\mu$~\cite{SRT,SRT2}.
Hereafter, the term of projective geometry is still used in this
sense.

If the \PoRcl\ is regarded as the fundamental principle for the
physics at cosmic scale, it is needed to answer the questions: How
to set up these inertial frames? And  how to face Einstein's
``argument in a circle" for the \PoR?~\cite{1922}. With the help of
{the evolution of our universe}, the inertial frames can definitely
be set up  via the extremely asymptotic behavior of the universe.
Actually, the time arrow of the universe should coincide with the
cosmic time of the Robertson-Walker-like \dS\ cosmos that is related
to the Beltrami inertial frames on the Beltrami-\dS\ spacetime by
changing the propertime simultaneity to the Beltrami coordinate time
and vice versa. Thus, the evolution of our universe can definitely
mark the time axis of Beltrami inertial frames so does for  all
kinds of  inertial frames~\cite{PoI}. This is completely different
from Einstein's manner for determining the inertial frames with ``
an argument in a circle"~\cite{1922}. As for Einstein's question:
``Are there at all any inertial systems for very extended portions
of the space-time continuum, or, indeed, for the whole
universe?"~\cite{1922} Our answer in positive, otherwise it is
impossible to set up kinematics at cosmic scale based on the \PoRcl.
We will explain this  later.

\omits{ and how to answer Einstein's question: ``Are there at all any
inertial systems for very extended portions of the space-time
continuum, or, indeed, for the whole universe?"\cite{1922} In fact, via its asymptotic
behavior the evolution of the universe can definitely
mark the time-axis of Beltrami inertial frames so that other kinds of inertial motion can also be fixed by the time-arrow of the universe and so does for inertial frames in general. This way of determination of inertial frames is completely different from Einstein and is away from the ``argument in a
circle"\cite{PoI}.}

As was shown recently, based on the
 \PoRcl, with common Lorentz isotropy for the \Mink-lightcone $C_O$ with its space/time-like region $R_\pm$ at the origin (see Eqs. (\ref{eq:CO}) and (\ref{eq:Rpm})),
 there is an \SR\ triple~\cite{SRT,SRT2,SRT3} for
  three kinds of \SR\ of Poincar\'e/\dS/\AdS\ group
${\cal P}/{\cal D}_\pm$ invariance on \Mink/Beltrami-\dS/\AdS\
spacetime $M/D_\pm$, respectively~\cite{Lu, LZG, Italy, BdS,TdS,
PoI, dual07,Lu80, Yan}. In addition, there also exists a dual
Poincar\'e group ${\cal P}_2$ of \LFT s that include the projective
infinity. So, they turn  the lightcone (\ref{eq:CO}) with its
regions (\ref{eq:Rpm}) to the compact one $\bar C_O$  with its
space/time-like domains $\dot R_\pm$  and also preserve them as a
compact origin lightcone structure $[\bar C_O]$  
in (\ref{eq:CObar}) and (\ref{eq:Rpmdot}). %
Although the formulae are almost the same, the meaning is quite
different. For the domain $\dot R_\pm$, there is  some (projective)
infinity from the \LFT s of the dual Poincar\'e group. And  with
boundary as the compact lightcone $\bar C_O$,  $\dot R_\pm$ is
associated to  region $R_\pm$ (\ref{eq:Rpm}) at the common origin in
\Mink/\dS/\AdS-space{time} $M/D_+/D_-$, respectively~\cite{SRT3}.
Further, there is also a pair of ${\cal P}_2$-invariant degenerate
Einstein manifolds $M_\pm:=(M^{{\cal P}_2}_\pm, {\vect g}_\pm,
{\vect g}^{-1}_\pm, { \nabla_{\Gamma_\pm}})$ of $\Lambda_\pm=\pm 3
l^{-2}$ for the domain $\dot R_\pm$ induced from $\bar
C_O=\partial\dot R_\pm$ as an absolute in the \PDG\
approach~\cite{SRT3}. Since for $M_\pm$  the compact lightcone $\bar
C_O$  with  the origin is excluded, the dual Poincar\'e group ${\cal
P}_2$  cannot offer independently a meaningful physical 4d
kinematics, but as  a kinematical symmetry it still is associated to
three kinds of \SR\ as a triple and form a relativistic quadruple
together with the Poincar\'e/\dS/\AdS\ groups and their invariant
spacetimes.

 As  subalgebras of 
 $\frak{im}(1,3)\cong \frak{pgl}(5,R)$ with  common
$\frak{so}(1,3)$, the \dS/\AdS\ algebras $\frak{d}_\pm:=\frak{so}(1,4)/\frak{so}(2,3)$ and the
Poincar\'e/dual  Poincar\'e algebras
 $\frak{p}/\frak{p}_2$ are related
 by the combinatory relation, which will be shown in (\ref{eq:combinatory}), \cite{SRT, SRT2} %
 among their translations $(\P_\mu, \P'_\mu,\P^\pm_\mu)$ including the pseudo ones $\P'_\mu$
 as representations of Lorentz algebra, respectively. Thus, with common ${\frak so}(1,3)$,
 each two of them construct an algebraic doublet,
    so there are six doublets in total. All of them  form an algebraic quadruplet $\frak
q:=(\frak{p},\frak{p}_2, \frak{d}_+,\frak{d}_-)$. 
This is not only different from the
 contraction \cite{IW,Bacry} and deformation~\cite{deform} approaches, but also very different from conventional concept on the
  inertial frames in certain given spacetime geometry. This relation, in fact, indicates that different kinds of inertial frames of
 ${\cal P}/{\cal P}_2/{\cal D}_\pm$-invariance can be described uniformly by the same inhomogeneous projective coordinates
  and their differences can be marked by the corresponding
  translations.

  Actually, corresponding to the algebraic relation (\ref{eq:combinatory}),
 four groups are also related 
 with common Lorentz isotropy ${\cal L}:=SO(1,3)={\cal
P}\cap{\cal
 P}_2\cap
{\cal D}_+ \cap{\cal D}_-$. And for the Poincar\'e/\dS/\AdS\ group
${\cal P}/{\cal D}_+/{\cal D}_-$, their invariant spacetimes share
the common origin lightcone structure $[C_O]$ shown in
(\ref{eq:[Co]}) with the same space/time-like region $R_\pm= M\cap
D_+\cap D_-$. While, for the lightcone $C_O$ and its region $R_\pm$,
the \LFT s of dual Poincar\'e group generate the ${\cal
P}_2$-invariant compact lightcone $\bar C_O$ with infinity included
and its space/time-like domain $\dot R_\pm$, $\partial \dot
R_\pm=\bar C_O$.  And induced from the compact origin lightcone
$\bar C_O$ as the absolute, there is an Einstein's manifold
$M_\pm$ for the  domain $\dot R_\pm$, respectively. Then, with the
lightcone structure $[C_O]$ or its  ${\cal P}_2$-invariant compact
partner $[\bar C_O]$ (\ref{eq:CObar}), 
there are 
six doubles
$[{\cal P},{\cal
  P}_2]_{M/M_\pm}$,  $[{\cal D}_\pm,{\cal P}]_{D_\pm/M}$,
 $[{\cal D}_\pm,{\cal P}_2]_{D_\pm/M_\pm}$ and $[{\cal D}_+,{\cal D}_-]_{D_\pm}$. They
 form a relativistic  quadruple
${\frak Q}_{PoR}=[{\cal P}, {\cal P}_2, {\cal D}_+,{\cal
D}_-]_{M/M_\pm/D_\pm}$. It is associated to the triple of three
kinds of \SR\ on 
\Mink/\dS/\AdS-spacetimes~\cite{SRT3}.

In fact, the geometry of $M/M_\pm/D_\pm$ with ${\cal P}/{\cal
P}_2/{\cal D}_\pm$-invariance is the subset of 4d projective
geometry invariant under  $ PGL(5,R)\sim IM(1,3)$, respectively. In
\cite{SRT3}, the transformations of the groups ${\cal D}_+/{\cal
D}_-/{\cal P}_2$ and their geometric properties on the
Beltrami-\dS/\AdS\ space{time} $D_\pm$ and the degenerated Einstein
manifold $M_\pm$ for domain $\dot R_\pm$ have been studied by the
projective geometry method, respectively. For instance, regarding
the boundary of the domain for the Beltrami-\dS/\AdS\  spacetime as
an absolute shown in (\ref{eq:domains}) and (\ref{eq: sigma=0}),
 respectively,
 the corresponding \LFT s of \dS/\AdS\ group can be obtained from the \LFT s of  group $IM(1,3)$ to
 preserve  them.
Then the metric in the domain and other geometric properties
invariant under the \dS/\AdS\ \LFT s can also be gotten from the cross ratio invariant.

 However, the \dS\ and \AdS\ space{time}s may have different radii. And
the realistic cosmological constant  in the universe may  contain
two parts: an invariant part as a fundamental constant and a slightly
variable part. In order to describe these properties, some
variable numbers may be introduced. On the other hand, it is easy to find that
the corresponding formulae in different cases are very similar to each other.
Actually, as long as the domain 
$r_\pm(x)$ in (\ref{eq:Rpm}) for $\dot R_\pm$ is replaced by
the domain $\sigma_\pm(x)$ (\ref{eq:domains}) for the
Beltrami-\dS/\AdS\ spacetime, the metric, Christoffel connection,
Riemann curvature and so on of the degenerated Einstein manifold
$M_\pm$ with the cosmological constant $\Lambda_\pm=\pm 3 l^{-2}$
become that of the Beltrami-\dS/\AdS\ spacetime as the
positive/negative constant curvature Einstein manifold with
$\Lambda_\pm=\pm 3 l^{-2}$ and vice versa, respectively. These
indicate that all these spacetimes and their invariant groups 
may be dealt with in a uniform manner at same time based on the
\PoRcl\ no matter what  the \dS/\AdS\ radius should be and whether
the $\Lambda$ is variable.

We will show in this paper that these can all be reached by what is
named the Poincar\'e-\dS\ flow ${\cal F}_{{r},{s}}$. It is just  a
parameterized quadruple with a pair of real numbers $\{{r},{s}\}$
for the domain, absolute, group transformations  and other
properties.
For four different regions of these numbers, the flow corresponds 
to
four cases, respectively. Thus, to deal with  the flow
is just to study all the members in the quadruple simultaneously. Actually,
in the flow the universal constant 
   $l$ is  replaced by $l|s|^{-1/2}$ in relevant cases, so the
 \dS/\AdS\ spacetimes of variable  radius $l|s|^{-1/2}$ appear. Correspondingly,  the value of cosmological constant
 $\Lambda_s=3sl^{-2}$ is also  changed for different $s$. It may also allow the other
 number $\{r\}$ to change slightly at above fixed values, too, especially
 for the case of dual Poincar\'e group.
 For usual Poincar\'e group, the first region shown in Eqs. (\ref{eq:tsf}) and (\ref{eq:ts})
  makes sense only for
   on the hyperplane at infinity in 4d real projective space. It is the absolute invariant under
   usual Poincar\'e group.

In terms of the  projective geometry method, especially Hua's matrix geometry
 (see, \textit{e.g.} \cite{Hua, HuaC, Hua63, HuaWan}), from the
invariance of Poincar\'e-\dS\ absolute (\ref{eq: sigmats=0}),  it
follows the Poincar\'e-\dS\ \LFT s  of the flow reduced from \LFT s
of $IM(1,3)$ (\ref{eq:LFT}). And from the cross ratio invariant, the
Beltrami metric and other properties follow. Further, Newton's law of inertia and the
$1+3$ decomposition with respect to  Beltrami coordinate
simultaneity can be given uniformly. The latter shows that the
coordinate simultaneous hyper-surfaces are differomorphic to 3d
Euclid{ean} space, sphere or pseudo-sphere for suitable values of
$\{r,s\}$, respectively. In fact, for four regions of $\{r,s\}$
shown in (\ref{eq:ts}), the Beltrami
 metric  of  the flow
  turns to the \Mink-metric on \Mink-space{time} $M$, the ${\cal P}_2$-invariant metric of degenerated Einstein
  manifold $M_\pm$ for $\dot R_\pm$ with $\Lambda_s=3s l^{-2}$,
   whose  coordinate time is just the degenerated direction, %
  and the Beltrami metric on the Beltrami-\dS/\AdS-space{time} $D_\pm$ of radius $l|s|^{-1/2}$, respectively.

As in the case of the Beltrami-\dS/\AdS\ spacetimes, there is also
another simultaneity in the
 flow, \textit{i.e.} the propertime simultaneity. The
Robertson-Walker-like cosmos of the flow with different space
foliations can also be studied and it coincides the cosmological
principle with  Poincar\'e-\dS\ invariance. Thus, the
Poincar\'e-\dS\ flow is out of the puzzle between the \PoR\ and the
cosmological principle in Einstein's theory of relativity, which had
been emphasized by Bondi~\cite{Bondi}, Bergmann~\cite{Bergmann} and
Rosen~\cite{Rosen} long ago and implied by Coleman,
Glashow~\cite{CG} and others recently. For the case of $r=1, s\in
R^+$, the flow turns to the \dS\ space{time} with  radius $l
s^{-1/2}$. It is clear that the \dS\ space{time} can  fit
kinematically at the accelerated expanding phase with an entropy
upper bound (\ref{S_R}) for the evolution of the universe. Moreover,
it is also away from Einstein's ``argument in a circle" for the
\PoR~\cite{1922}, since the time-arrow of the universe can indicate
the existence of the Beltrami-frame of inertia via its
Robertson-Walker-like counterpart whose time axis  should coincide
with the cosmic time-arrow of the universe~\cite{PoI}. This also
proves the existence of the inertial frames in the universe. In
addition, as different $s$ may be chosen, it may also provide an
inflationary phase near the Planck scale $\ell_P$ with an entropy
bound from below for the beginning of the universe or the one at
other scales like the GUT scale for the inflation. Although the
latter is different from the inflation model, its role to the
inflation should be studied further.

Since the Poincar\'e-\dS\ flow is based on the \PoRcl, it is away from the framework
of general relativity. So, there is no gravity for the spacetime of the flow as an Einstein manifold of
$\Lambda_s$, which contains all spacetimes $M/M_\pm/D_\pm$. 

The paper is arranged as follows.  In sec. \ref{sec:PoR}, we briefly
recall the \PoRcl, the inertial motion group and the relativistic
kinematics that act as the \SR\ triple associated with a
relativistic quadruple of four types of related  groups and
geometries. In sec. \ref{sec:PDflow}, we introduce the
Poincar\'e-\dS\ flow and  deal with the symmetry and geometry
aspects as well as the physical issues for inertial motions
uniformly  by means of the \PDG\ method and corresponding embedding
approach. In sec. \ref{sec:Cosm}, we study the cosmological meaning
of the flow. Especially, we show that the \dS\ \SR\ with the
\dS-dual Poincar\'e double $[{\cal D}_+,{\cal P}_2]_{D_+/M_+}$ can
provide the consistent kinematics for  physics at the cosmic scale.
It may also work near the Planck length or at other scales. Finally,
we  end with some remarks.
\section{The $\boldsymbol{PoR_{cl}}$, Inertial Motion Group and Relativistic
Quadruple}\label{sec:PoR}%

\subsection{The $\boldsymbol{PoR_{cl}}$ and  Inertial Motion Group}\label{sec: UWFH}

Long ago, Lu~\cite{Lu}  suggested that Einstein's 
\SR\ \cite{1905, 1922} should be extended to \dS/\AdS\ space{time} of
constant curvature, and studied the \dS/\AdS\ invariant \SR\
\cite{Lu,LZG}\footnote{Similar consideration had also been made by
Fanttappi\'e \textit{et al} earlier in view of Erlangen
program rather than the \PoR\ (see, \textit{e.g.} \cite{Italy}).}.
Recently, motivated by precise cosmology, much studies have been
made \cite{BdS, dual07}. In fact, if the Euclidean assumption on
space is avoided, the inertial motion of Newton's first law can
reach the infinity. In fact, the kinematic aspect of Newton's first
law in inertial coordinate systems can be considered first in
general  and it follows the inertial motion group $IM(1,3)\sim
PGL(5,R)$ with algebra $\frak{im}(1,3)\cong \frak{pgl}(5,R)$
kinematically \cite{SRT,SRT2}.

Umov~\cite{Umov} , Weyl~\cite{Weyl}, Fock~\cite{Fock} and
Hua~\cite{Hua, HuaC} studied the important issue long ago: What are
the most general transformations among the inertial frames ${\cal
F}:=\{S(x)\}$ that keep the inertial motions?

Based on the \PoRcl, the  answer can be stated as: {\it The most
general transformations among ${\cal F}$ that preserve
 inertial
motion \omits{in the frames ${\cal F}$\footnote{ Conventionally,
the concepts are depended on certain metric introduced. These had
been generalized in \cite{SRT, SRT2, Umov, Weyl, Fock}.}}
 \be\label{eq:uvm}%
 x^i=x_0^i+v^i(t-t_0),~~
v^i=\frac{dx^i}{dt}={\rm consts}.~~ i=1, 2, 3, %
\ee%
where  the isotropy among space coordinates is  required, are  the  \LFT s    of twenty-four parameters %
\be\label{eq:LFT} %
T:\quad&& l^{-1}x'^\mu = \frac{A^\mu_{\ \nu} l^{-1} x^\nu +
b^\mu}{c_\lambda l^{-1}x^\lambda + d},\quad x^0=ct,
\\\label{eq:det}
 && det\ { T}=\left| \begin{array}{rrcrr}
    A    & b^t \\
    c
     & d
  \end{array} \right|
  \neq 0,
\ee%
where $A=(A^\mu _{~\nu})$  a $4\times 4$ matrix, $b,c$ $1\times 4$
 matrixes, $d\in R$, $c_\lambda x^\lambda=\eta_{\lambda\sigma}c^\lambda
 x^\sigma$,
  $\small{^ t}$
 for  transpose and  $J=(\eta_{\mu\nu})=diag(1,
  -1,-1,-1)$. These $LFTs$
  form the inertial motion
 group $IM(1,3)$ homomorphic to  the 4d real projective
group ${PGL}(5,R)$ with the inertial motion algebra
$\mathfrak{im}(1,3)$ isomorphic to $\frak{pgl}(5,R)$.} Further,
the time reversal ${T}$ and space inversion ${P}$ preserve the
inertial motion (\ref{eq:uvm}). So, all  issues are modulo the $T$
and ${P}$ invariance.

Inertial motion (\ref{eq:uvm}) can be viewed as a 4d
 straightline with an affine parameter $\lambda$%
 \be\label{eq:affine line}%
 x^\mu(\lambda)=x^\mu_0+\lambda \upsilon^\mu, \quad \upsilon^\mu=consts,%
 \ee%
where $-\infty<\lambda<\infty$. If the inertial motion may reach the
infinity so the infinity should be included, it becomes a projective
straightline~\cite{HuaC, Hua}.
Then from the fundamental theorem in projective geometry
\cite{HuaWan}, it follows the \LFT s in (\ref{eq:LFT}) that
transform some events \{$X(x^\mu):\, c_\lambda l^{-1}
x^\lambda+d=0$\} to the infinity.  And \LFT s (\ref{eq:LFT}) form
the real projective group ${PGL}(5,R)$ with algebra
$\frak{pgl}(5,R)$. But, for orientation in physics the antipodal
identification for $x^\mu$ as inhomogeneous projective coordinates
of 4d real projective space should not be taken so that
$IM(1,3)\thicksim{PGL}(5,R)$. As was mentioned, we still call it the
projective geometry approach in this sense, hereafter.

The set
 $\{T\}^\mathfrak{im}:=(\H^\pm, \P^\pm_i,\J_i,\K_i,\N_i,\R_{ij}, M_\mu)$ of generators
 for \LFT s (\ref{eq:LFT}) spans the
 $\frak{im}(1,3)$
 \cite{SRT2},
where
 \be\nno%
 &\N_i := t \partial_i + c^{-2} x_i
\partial_{t},&\\\label{eq:R,M}
&R_{ij}:= x_i \partial_j + x_j \partial_i, (i< j),\,\, M_\mu :=
x^{(\mu)} \r_{(\mu)},&
\ee%
where no summation 
for  repeated  indexes in brackets. The Lorentz isotropy algebra
$\frak{so}(1,3)$ of Lorentz group
${\cal L}$ generated by  space rotation ${\bf J}_i$ defined as%
  \be\label{eq:J}%
 \J_i=\frac{1}{2}\epsilon_i^{\,jk}L_{jk}, \ \, L_{jk}:=x_j\partial_k-x_k\partial_j,%
 \ee%
  and Lorentz
boosts $\K_i$ defined as%
\be\label{eq:Ki}%
 \K_i:=t \partial_i -c^{-2} x_i
 \partial_t.%
\ee%
Among four  time and space
translations $\{{\cal H}\}:=\{ \H, \H', H^\pm\}$ of dimension
$[\nu]$,  $\{{\bf P}\}:=\{ {\bf P}_i, {\bf P}'_i, {\bf P}^\pm_i\}$
of dimension $[l^{-1}]$ including the pseudo-ones $(H',\P'_i)$, and
four boosts $\{ \K\}:=\{\K_i, \N_i,
 \K^{\mathfrak{g}}_i,\K^{\mathfrak{c}}_i\}$ of Lorentz, 
 geometry, 
 Galilei 
  and Carroll boosts of dimension $[c^{-1}]$,  there are two
 independences, respectively
  \be \label{eq:H}
&H:=\partial_t,\,\, H':=-\nu^{2}t x^\nu\partial_\nu,\,\, H^\pm
:=\partial_t\mp \nu^{2}t x^\nu\partial_\nu;&\\\label{eq:P}
 &\P_i:=\partial_i,\, \P'_i:=- l^{-2}x_i
 x^\nu\partial_\nu,\, {\mathbf P}^\pm_i :=\partial_i\mp l^{-2}x_i
 x^\nu\partial_\nu;&\\\nno%
& \K_i:=t \partial_i -c^{-2} x_i
 \partial_t,\,\,\, \N_i:=t \partial_i +c^{-2} x_i
 \partial_t,&\\\label{eq:K}& \K_i^\frak{g}:=t \partial_i,\quad \K_i^\frak{c}:= -c^{-2} x_i
 \partial_t,&
\ee%
where $\nu:=c/l$ is called the Newton-Hooke constant. These
generators are scalar and vector representation of $\frak{so}(3)$
generated by space
  rotation  ${\bf J}_i$ without dimension
 as follows~\cite{SRT, SRT2}
\be\label{eq: JHPK}
[ \J,\J ]=\J, \,\,[\J, {\cal H}]=0,\,\, [\J, {\P}]={\P},
[\J, {\K}]={\K},
\ee%
where with
$\epsilon_{123}=-\eps_{12}^{\ \ 3}=1$ 
 $[\J,\P]=\P$ is, \textit{e.g.} a shorthand 
of $[\J_i,\P^\pm_j]=-\epsilon_{ij}^{~~k}\P^\pm_k$ etc.  All
generators and commutators  have right dimensions expressed by the
constants $c, l$ or $\nu$.

It is important that there are
combinatory relations among those translations and boosts %
\be\label{eq:combinatory}
&H\pm H'=\H^\pm, \quad \P_i\pm \P'_i=\P_i^\pm,&\\\label{eq:combinatory2}%
&\K_j/\N_j=\K_i^\frak{g}\pm \K_i^\frak{c}.& %
\ee%
These relations mean that for different kinematics with corresponding translations and boost for different spacetimes as well as spaces and times the inhomogeneous projective coordinates $x^\mu$ do make sense as inertial coordinates. This is completely different from usual understanding in conventional approach. 

In Table I, all relativistic kinematics are listed symbolically. For
the geometrical and non-relativistic cases~\cite{SRT2}, we shall
study them in detail elsewhere. {\bf\begin{table}[thp] \caption{ All
relativistic kinematics}\vskip 2mm {\small
\begin{tabular}{|c c c c c c c c|}
\hline Group & Algebra & Generator Set
  & $[\cal H,\P]$ & $[\cal H,\K]$ & $[\P,\P]$ &$[\K,\K]$ &$[\P,\K]$\\
\hline ${\cal D}_+$  &$\mathfrak{d}_+$ & $(H^+, \P^+_i, \K_i,
\J_i)$ & $\nu^2\K$ & $ \P$ & $l^{-2}\J$ &
$-c^{-2}\J$  &$c^{-2}\cal H$\\
${\cal D}_-$ & $\mathfrak{d}_-$ &  $(H^-, \P^-_i, \K_i, \J_i)$ &
$-\nu^2\K $ & $ \P$ & $-l^{-2}\J $
& $-c^{-2}\J$ & $c^{-2}\cal H$\\
$\begin{array}{c}{\cal P}\\{\cal P}_2\end{array}$ &$\begin{array}{c} \mathfrak{p}\\
\mathfrak{p}_2\end{array}
$ & $\begin{array}{c}(H, \P_i, \K_i, \J_i)\\
\{H', \P'_i, \K_i, \J_i\}\end{array}$ & 0 & $\P$ & 0 &$-c^{-2}\J$&$c^{-2}\cal H$\\
\hline
\end{tabular}
}

\end{table}}

It is clear that
with common $\frak{so}(1, 3)$ isotropy, the Poincar\'e algebraic
doublet $(\frak{p},\frak{p}_2)$ leads to  the \dS/\AdS\ algebraic
doublet $(\frak{d}_+,\frak{d}_-)$  with the Beltrami time and
space translations $(\H^\pm, \P^\pm_j)$ in the
Beltrami-\dS/\AdS-space{time}, respectively, and vice
 versa~\cite{SRT, SRT2}. In addition,
from the algebraic relations  of $\frak{im}(1,3)$ \cite{SRT2} both the Poincar\'e doublet
$(\frak{p},\frak{p}_2)$ and the \dS/\AdS\ doublet $(\frak{d}_+,\frak{d}_-)$ are closed
in the  $\frak{im}(1,3)$,  while the  generators
 $(R_{ij}, M_\mu)$, which generate $A_{\,\,\nu}^\mu$ in \LFT s (\ref{eq:LFT}) together
 with $\N_i$ in (\ref{eq:K}) and
 $(\K_i, \J_i)$, 
 exchange the  translations in the \dS/\AdS\ doublet $(\frak{d}_+,\frak{d}_-)$ and keep that in the Poincar\'e doublet
 $(\frak{p},\frak{p}_2)$ \cite{SRT,SRT2}. Since all issues are in  $IM(1,3)$ based on the
\PoRcl,  there should be
three kinds of \SR\ %
that form {\it the \SR\ triple} of four related groups   ${\cal
L}={\cal P}\cap{\cal
 P}_2\cap
{\cal D}_+ \cap{\cal D}_-$ as a relativistic quadruple ${\frak
Q}_{PoR}:=[{\cal P}, {\cal P}_2, {\cal D}_+,{\cal
D}_-]_{M/M_\pm/D_\pm}$~\cite{SRT3} with the algebraic quadruplet
$\frak{q}=(\frak p,\frak{p}_2, \frak{d}_+,\frak{d}_-)$~\cite{SRT2}.
\omits{This also the case for their relevant groups and geometries.} %


\subsection{The Poincar\'e Double, Special Relativity Triple and Relativistic Quadruple}


It is clear that for $A=L\in \cal L$, $c=0$ and $d=1$, the \LFT s in
(\ref{eq:LFT}) reduce to the transformations of usual Poincar\'e
group
 ${\cal P}:=ISO(1,3)=R(1,3)\rtimes \cal L$
\be\label{eq:ISO(1,3)}%
 {P}: \,\,
 x'^\mu=
 (x^\nu-a^\mu)L^\mu_{\ \nu},\,\, det{P}= det \left (
\begin{array}{rrcrr}
    L & b^t \\
    0 & 1
  \end{array} \right)
  \neq 0,%
\ee%
where $R(1,3)$ is 4d translation group with respect to parameters $a^\mu$ and $L=(L^\mu_{\ \nu})\in  SO(1,3)$,  $ b^t=-l^{-1}(aL)^t$. And
$\forall P\in {\cal P}$, 
it follows the \Mink-space{time} $M:={\cal P}/{\cal L}$ with  the
\Mink-metric and the \Mink-lightcone
at  event $A(a^\mu)$ %
\be\nno%
&ds^2=\eta_{\mu\nu}dx^\mu dx^\nu=dxJdx^t,&\\\label{eq:MconeatA}
&C_A:\,\,\,
(x-a)J(x-a)^t=0,&  
\ee%
with the lightcone structure $[C_A]$ 
:%
 \be\label{eq:CA}
[C_A]:\,\, (x-a)J(x-a)^t \gtreqless0,\,\,ds^2|_A=dxJdx^t|_A=0.
\ee%
For the origin lightcone $C_O$ and its space/time-like region $R_\pm$, we have%
  \be\label{eq:CO}%
&C_O: \,\, \eta_{\mu\nu}x^\mu x^\nu= xJx^t=0, &\\\label{eq:Rpm}%
&R_\pm: \quad xJx^t\lessgtr 0;&\\\label{eq:[Co]}
&[C_O]:\,\, xJx^t \gtreqless0,\quad ds^2|_O=dxJdx^t|_O=0.& \ee%

The set $\{T\}^{\mathfrak{p}}:=(H,\P_i, \K_i,\J_i)$ spans 
$\frak{p}:=\mathfrak{iso}(1,3)$ 
listed in Table I symbolically. As was just mentioned, in view of
\PDG, Poincar\'e group keeps a hyperplane at infinity as its
absolute in 4d real projective space $RP^4$. In terms of
homogeneous projective coordinates $\xi^\mu, \xi^4$, if $\xi^4\neq
0$, the \Mink-coordinates $x^\mu$ may be regarded as
\be\label{eq:xi4n0}%
x^\mu=l{\xi^4}^{-1}{\xi^\mu},\quad \xi^4 \neq 0,%
\ee%
while the hyperplane at infinity corresponds to%
\be\label{eq:xi420}%
\xi^4=0. \ee%

 It is important that  for $A=L\in \cal L$ and $b=0$ and $d=1$, the \LFT s in
(\ref{eq:LFT}) reduce to the  transformations 
\be\label{eq:LFTP2}
 l^{-1}x^\mu\to  l^{-1}x'^\mu = \frac{L^\mu_{\ \nu}l^{-1} x^\nu }{c_\lambda l^{-1} x^\lambda +
  d},
  \ee%
  in which 
for $d=1$ form a group called the dual
Poincar\'e group, $\forall P_2\in {\cal P}_2 \cong ISO(1,3)$, 
\be\label{eq:P2}%
 det P_2= det \left ( \begin{array}{rrcrr}
    L & 0 \\
    c & d
  \end{array} \right)
  \neq 0,\,\,\, {\rm for}\,\,\, d=1.&
\ee%
It is important that \LFT s (\ref{eq:LFTP2}) send those events $X(x^\mu)$ satisfying $c_\lambda l^{-1} x^\lambda +
  d=0$ to infinity. Therefore, corresponding infinity exists with the lightcone $C_O$ (\ref{eq:CO}) and its space/time-like region $R_\pm$ (\ref{eq:Rpm}) and the dual Poincar\'e group turn them to the compact one $\bar C_O$ 
  and its 
  space/time-like domain $\dot R_\pm$, and the former is the boundary of the latter, 
  $\bar C_O=\partial \dot R_\pm$, as follows%
  \be\label{eq:CObar}%
&\bar C_O:\,\,xJx^t|_{{\cal P}_2}=0,&\\\label{eq:Rpmdot}%
&\dot R_\pm: \quad r_\pm(x):=\mp xJx^t|_{{\cal P}_2}> 0,\quad \partial \dot R_\pm=\bar C_O.& \ee%
Clearly, \LFT s (\ref{eq:LFTP2}) do preserve the compact lightcone structure consists of (\ref{eq:CObar}) and (\ref{eq:Rpmdot}). Hence,  the symmetry of them
 is not just the Lorentz group,
but the semi-product of the dual  Poincar\'e group ${\cal P}_2$ and
a dilation.

 In fact,  in terms of the \PDG, from
 $\bar C_O$ (\ref{eq:CObar}) and $\dot R_\pm$ (\ref{eq:Rpmdot})  as   absolute and  domain, it follows that \LFT s  (\ref{eq:LFT}) are reduced to their
   subset  (\ref{eq:LFTP2}). 
    And the set $\{T\}^{\frak{p}_2}=
 (\H',
 \P'_i,\K_i,\J_i)$
 spans
 $\frak{p}_2 \cong\frak{iso}(1,3)$  and  the generator of dilation  $d$ in (\ref{eq:LFTP2}) is
$D=x^\mu\partial_\mu=\sum M_\mu\in {\frak im}(1,3)$.

Further, for a pair of events $A(a^\mu), B(b^\mu)\in \dot R_\pm$,  a
line between them%
\be\label{eq:L}%
(1-\tau)a+\tau b %
\ee%
 crosses the absolute $\bar{C}_O=\partial(\dot R_\pm)$ at $\tau_1$,
$\tau_2$. For four events with $\tau=(\tau_1, 1, \tau_2, 0)$, a
cross ratio
can be given. From the power-2 cross ratio invariant%
\be\label{eq:Deltapm}%
&{\Delta}_{R_\pm}^2(A, X)
=\pm l^{2}\{r_\pm^{-1}(a)r_\pm^{-1}(x)r_\pm^2(a,x)-1\},&\\\nno
&r_\pm(a):=r_\pm(a,a)>0,\,\ r_\pm(a,x):=\mp aJx^t,& %
\ee%
 for two closely nearby  events
$X(x^\mu), X+dX(x^\mu+dx^\mu)\in \dot R_\pm$, it follows a  metric
\be\nno%
&{g}_{\pm \mu\nu}= 
( \frac{\eta_{\mu\nu}}{r_\pm(x)}\pm
 l^{-2}\frac{x_\mu
x_\nu}{r^2_\pm(x)}),\,\ 
\,x_\mu=\eta_{\mu\lambda}x^\lambda,&\\ \label{eq:gMpm}%
&{\vect g}_\pm:=({ g}_\pm)_{\mu\nu}=
{r^{-1}_\pm(x)} ( J \pm l^{-2} \d {J x^t x J }{r_\pm(x)}).
\ee%
Although it is degenerated, \textit{i.e.} $det\,{\vect g}_\pm=0$,
formally there is still a contra-variant metric as its inverse,
respectively: %
\be\label{eq:hMpm}%
&{\vect g}^{-1}_\pm(r_\pm)= 
{r_\pm(x)} ( J
\pm  l^{-2}\d {J x^t x J }{r_\pm(x)})^{-1},&%
\ee%
which is divergent with a finite part
\begin{equation}
{\vect g}^{-1}_\pm(r_\pm)\big|_{finite}=l^{-2}r_\pm J x^t x J.
\end{equation}

The Christoffel
symbol for dual Poincar\'e symmetry can still be formally obtained 
 \be\label{eq:ChrsMpm}%
 \Gamma^\lambda_{\pm
\mu\nu} (x)= \pm r_{\pm}^{-1}(x) ( \delta^\lambda_\mu x_\nu +
\delta^\lambda_\nu x_\mu).%
 \ee%
It is obviously metric compatible, \textit{i.e.} %
$\nabla_{\Gamma_\pm}{\vect g}_\pm=0,
\,\,\,\,\nabla_{\Gamma_\pm}{\vect g}^{-1}_\pm=0, \,\,\,\,\nabla_{\Gamma_\pm}{\vect g}^{-1}_\pm\big|_{finite}=0$. %
Further,  the Riemann, 
Ricci
  and scale curvature reads, respectively:
 \be\nno
&R^\mu_{\pm \nu \la \si}(x)=\pm
l^{-2}(g_{\pm\nu\la}\dl^\mu_\si-g_{\pm\nu\si}\dl^\mu_\la)&\\\label{eq:curMpm}
&R_{\pm \mu \nu }(x)
= \pm 3 l^{-2}g^\pm_{\mu\nu}, \,~~\,R_\pm(x) = \pm  12 l^{-2}.&%
\ee %
So, the  $M_\pm:=(M^{{\cal P}_2}_\pm, {\vect g}_\pm,
{\vect g}^{-1}_\pm, { \nabla_{\Gamma_\pm}})$ 
is an Einstein
 manifold with 
 $\Lambda_\pm=\pm 3/l^2$ for $\dot R_\pm$  (\ref{eq:Rpmdot}), respectively.
In \cite{Huang0909},
 a different ${\cal P}_2$-invariant kinematics with degenerated geometry for $R_+$
 is given in another  manner. 

 Since all these  objects are given originally from
 the cross ratio invariance of $\bar C_O$ (\ref{eq:CObar}) under \LFT s (\ref{eq:LFT}),
 such a ${\cal P}_2$-invariant kinematics is
 invariant under  (\ref{eq:P2}) and is based on the \PoRcl.  It easy to check
 that their Lie derivatives vanish with respect to  the ${\mathfrak p}_2$-generators
 as Killing vectors on  $M_\pm$,
 respectively, and that Eq. (\ref{eq:uvm}) is indeed the
  equation of motion
 for a free particle, if any, on such a pair of Einstein manifolds
 $M_\pm$.

It is clear that in all the formulae the origin and the compact
lightcone $\bar C_O$ are automatically excluded. For the origin
excluded, the name of dual Poincar\'e group makes sense dual to
usual Poincar\'e group, for which  the infinity is excluded. Due to
the origin is excluded, the ${\cal P}_2$-invariant kinematics is not
independently a meaningful physical one, rather as a kinematical
symmetry, which is always associated to three kinds of \SR\ as a
triple.

In fact, the above ${\cal P}_2$-structure also exists  with respect to the lightcone structure  $[C_A]$ (\ref{eq:CA}),  
while  the ${\cal P}_2$-structure with respect to the  $[C_O]$ 
 is just a
representative. This can be easily seen from all above relevant
formulae if all $x^\mu$ are replaced by $x^\mu-a^\mu$.   There are
intersections for ${\cal P}$ and ${\cal P}_2$, \textit{i.e.} ${\cal
L}= {\cal P}\cap{\cal P}_2$. While  for $M$ and $M_\pm$,  $R_\pm\in
M$ and $\dot R_\pm=\cap M_\pm$ are very closely related. Thus,
relevant to the \Mink-space{time} there exists a Poincar\'e double
$[{\cal P},{\cal P}_2]_{M/M_\pm}$ with a pair of Einstein
manifolds $M_\pm$ induced from $\bar C_O$ %
 for  $\dot R_\pm$, 
 respectively. Further, under usual Poincar\'e transformations
 (\ref{eq:ISO(1,3)}), 
 the
 Poincar\'e algebraic doublet $(\frak{p},\frak{p}_2)$ at the origin 
 can be
transformed to the event $A(a^\mu)$ and vice versa. In fact, under
transformations of usual Poincar\'e group $\cal P$, the generators
$\P'_\mu:=(H', \P'_i)$ as a 4-vector on $M$ are transferred and the
action is closed in the  algebra $\frak{im}(1,3)$ with
$\frak{d}_\pm$ for \dS/\AdS\ \SR\ as the subsymmetry:
\be\label{eq:[p,p']}%
 {\cal L}_{\P_\mu} \P'_\nu =[\P_\mu,\P'_\nu]\omits{=-l^{\-2}\eta_{\mu\nu}\sum_{\lambda}M_\lambda
 -l^{-2}\frac{1}{2}R_{\mu\nu}} \in \frak{im}(1,3)\cong
 \frak{pgl}(5,R).
 \ee%

Thus,  as long as the inertial motions for free particles and light
signals are naturally allowed to reach the infinity, the symmetry and
geometry related to the \Mink-space{time} are dramatically changed. In
order to eliminate the Poincar\'e double and restore Einstein's \SR, $\l =
\infty$ must be taken. This is equivalent to the Euclidean assumption
that  the inertial motions  cannot  reach the infinity.
However, the existence of the cosmological constant $\Lambda_+$ leads to a maximum value of $l$ and the Poincar\'e double  $[{\cal P},{\cal
P}_2]_{M/M_+}$ would appear with following bound   for  modern relativistic
physics%
\be\label{10-35}%
\nu^2:=(c/l)^2\simeq c^2\Lambda_+/3\sim
10^{-35}sec^{-2}.%
\ee%
 This value is so tiny that their
effects can still be ignored up to now locally
  except for the cosmic scale physics.


It is also shown \cite{SRT3} that   the \dS/\AdS\ double $[{\cal
D}_+,{\cal D}_-]_{D_\pm}$ of the algebraic doublet
$(\frak{d}_+,\frak{d}_-)$
 can  be reached  via the \Mink-lightcone  $C_O$ in (\ref{eq:CO})  by relaxing the flatness of \Mink-space{time}.
 In fact,
shifting  the  lightcone by  $\mp l^2$
 leads to the expression%
  \be\label{eq:domains}%
&\frak{D}_\pm:\quad \sigma_\pm(x):=\sigma_\pm(x, x)=1\mp l^{-2}xJ
x^t > 0,&%
\ee
which denote a pair of related regions on $M$ with boundary as a
`pseudosphere' 
\be\label{eq: sigma=0}%
&\frak{B}_\pm=\partial(\frak{D}_\pm):\quad  \sigma_\pm(x)=1\mp l^{-2}xJ x^t = 0,& %
 \ee
 respectively (see, \textit{e.g.} \cite{Synge}). %
 However, if the flatness of
 space{time} is relaxed,  regions (\ref{eq:domains}) and their boundaries
(\ref{eq: sigma=0}) without flatness are  just the the domain
conditions and their absolutes in the \PDG\ approach to  the
Beltrami model of \dS/\AdS-space{time} $D_\pm$ with radius $l$,
respectively~\cite{Hua, SRT3}. Actually, if  the coordinates $x^\mu$
are regarded as the Beltrami
coordinates in a chart $U_4$, say, 
\be\label{eq:BLxU4}%
x^\mu=l{\xi^4}^{-1}{\xi^\mu},\quad \xi^4> 0,%
\ee%
 regions  (\ref{eq:domains}) and boundaries (\ref{eq: sigma=0})
become the  \dS/\AdS-hyperboloid  ${\cal H}_\pm$ and their
boundaries,
respectively%
\be\label{eq:hyperboloids}%
 &{\cal H}_\pm:\quad \eta_{\mu \nu}\xi^\mu \xi^\nu \mp (\xi^4)^2 \lessgtr 0,&\\%
&{\cal B}_\pm=\partial{\cal H}_\pm:\quad \eta_{\mu \nu}\xi^\mu
\xi^\nu \mp
 (\xi^4)^2=0.&
\ee%
Clearly, the expressions of $\eta_{\mu \nu}\xi^\mu \xi^\nu$ and of
$(\xi^4)^2$ are in  intersections of the ${\cal H}_\pm$ and their
boundaries ${\cal B}_\pm=\partial{\cal H}_\pm$ and their vanishing expressions just correspond to the (compact) origin lightcone and hyperplane at infinity as the absolutes of the dual Poincar\'e group and of the usual Poincar\'e group, respectively.
Then, by means of the \PDG\ method, the intersected
Beltrami-\dS/\AdS-spacetimes $D_\pm$  invariant under the \dS/\AdS\ group ${\cal D}_\pm$  can
be set up  and form the \dS/\AdS\ double $[{\cal D}_+,{\cal
D}_-]_{D_\pm}$ with common Lorentz isotopy, \textit{i.e.} $\cal L={\cal D}_+\cap{\cal D}_-$. And they are closely related to  both the Poinca\'e group and the dual Poincar\'e group.

In fact,  \LFT s (\ref{eq:LFT}) with (\ref{eq: sigma=0}) as
the absolute  reduce to  the \dS/\AdS-\LFT s with  common
$L_{~\ka}^\mu\in\cal L$
\be\nno%
&{\cal D}_\pm:~ x^\mu\rightarrow {x'}^\mu=\pm
\sigma_\pm^{1/2}(a)\sigma_\pm^{-1}(a,x)(x^\nu-a^\nu)D_{\pm~\nu}^{~\mu},&\\\label{eq:SO14,23}
&D_{\pm~\nu}^{~\mu}=L_{~\nu}^\mu\pm l^{-2}%
a_\nu a^\ka (\sigma_\pm(a)+\sigma_\pm^{1/2}(a))^{-1}L_{~\ka}^\mu,&
\ee
which 
preserve  
(\ref{eq:domains}) for the Beltrami-\dS/\AdS\ space{time} $D_\pm$,
respectively. Further, from a power-2 cross ratio invariant, it
follows the interval between a pair of events $A(a^\mu)$ and
$X(x^\mu)$ and the lightcone with the vertex at $A(a^\mu)$ as
follows, respectively
 \be\label{eq:Dlt}%
 &{\Delta}_\pm^2(A,
X) =\pm
l^{-2}\{\sigma_\pm^{-1}(a)\sigma_\pm^{-1}(x)\sigma_\pm^2(a,x)-1\}\gtreqqless
0,
&\\\label{eq:Blightcone} %
&{\cal F}_{\pm}: \quad
\sigma_\pm^2(a,x) -\sigma_\pm(a)\sigma_\pm(x)=0.&%
\ee %
For the  closely nearby two events $X(x^\mu),
X+dX(x^\mu+dx^\mu)\in \frak D_\pm$,  the Beltrami metric ${\vect g}^B_\pm$ and its inverse  of \dS/\AdS\ space{time} follows
from (\ref{eq:Dlt}), respectively
\be\nno%
&ds_\pm^2=(\frac{\eta_{\mu\nu}}{\sigma_\pm(x)}\pm l^{-2}\frac{
x_\mu x_\nu}{
\sigma_\pm^{2}(x)})dx^\mu dx^\nu,\,\,\,\sigma_\pm(x)>0.&\\\label{eq: BKmetric}
&{\vect g}^B_\pm:=({ g}^B_\pm)_{\mu\nu}=
{\sigma^{-1}_\pm(x)} ( J \pm l^{-2} \d {J x^t x J }{\sigma_\pm(x)}).&\\\nno
&{\vect g}^{B -1}_\pm
={\sigma_\pm} ( J \pm l^{-2} \d {J x^t x J }{\sigma_\pm(x)})^{-1}=\sigma_\pm(J - l^{-2} x^t x).&
\ee%
Due to transitivity of (\ref{eq:SO14,23}), the Beltrami-\dS/\AdS\
space{time} ${D}_\pm \cong {\cal D}_\pm/\cal L$
 with $[C_O]=D_+\cap D_-$ is homogeneous,
respectively. It is also true for the entire Beltrami-\dS/\AdS\
space{time} globally.
 And the set
$\{T^{\frak{d}_\pm}\}=(\H^\pm, \P_i^\pm, \K_i, \J_i)$ of
(\ref{eq:SO14,23}) spans the \dS/\AdS\ algebra, respectively, and they form a doublet
$(\frak{d}_+,\frak{d}_-)$ with common Lorentz isotropy.

 In addition, the generators in 
$\{T^{\frak{d}_\pm}\}=(\H^\pm, \P_i^\pm, \K_i, \J_i)$ of the
\dS/\AdS\ algebra $\frak{d}_\pm$ can be regarded as Killing vectors
of Beltrami-\dS/\AdS\ metric (\ref{eq: BKmetric}). With respect to
these Killing vectors the  Lie derivatives vanish for the metric,
connection and curvature. Further, the geodesic motion of (\ref{eq:
BKmetric}) is indeed the inertial motion (\ref{eq:uvm}).

In short, domains $\frak{D}_\pm$ (\ref{eq:domains}) shifted from
$C_O$ (\ref{eq:CO}) indicate that the Beltrami-\dS/\AdS\ spacetimes $D_\pm$ share the same origin lightcone structure $[C_O]$ from
(\ref{eq:Blightcone}) and  (\ref{eq: BKmetric}). So, the \dS/\AdS\
\SR\ on $D_\pm$ form a  double $[{\cal D}_+, {\cal D}_-]_{D_\pm}$.
Since they also share the $[C_O]$ with \Mink-space{time} $M$, together
with  Poincar\'e,  \dS/\AdS-${\cal P}$, and  \dS/\AdS\ doubles,
there are also  \dS/\AdS-${\cal P}_2$ doubles $[{\cal D}_\pm,
{\cal P}_2]_{D_\pm/M_\pm}$ with Einstein manifold $M_\pm$ for  domain $\dot R_\pm$ (\ref{eq:Rpmdot}) generated by the ${\cal P}_2$-invariance. All six
doubles form the
 quadruple ${\frak
Q}_{PoR}$  with common Lorentz isotropy\omits{ and $R_\pm\supset M\cap
M_\pm\cap D_+\cap D_-$}. But, as far as the cosmological constants
$\Lambda_\pm$ are concerned, only two \dS/\AdS-${\cal P}_2$ doubles
$[{\cal D}_\pm, {\cal P}_2]_{D_\pm/M_\pm}$ are consistent
kinematically in principle.

\omits{As was mentioned, as long as domain condition $r_\pm(x)$
(\ref{eq:Rpm}) is replaced by domain condition $\sigma_\pm(x)$
(\ref{eq:domains}) in metric (\ref{eq:gMpm}), Christoffel connection
(\ref{eq:ChrsMpm}), Riemann curvature and so on in Eq.
(\ref{eq:curMpm}) of the degenerated Einstein manifold $M_\pm$ of
$\Lambda_\pm=\pm 3 l^{-2}$ becomes that of the Beltrami-\dS/\AdS\
space{time} as the positive/negative constant curvature Einstein manifold
with $\Lambda_\pm=\pm 3 l^{-2}$ and vice versa, respectively. This
may lead to the Poinca\'e-\dS\ flow  introduced in the next
section.}

\section{The Poincar\'e-$\boldsymbol{dS}$ Flow as a Parameterized Quadruple}\label{sec:PDflow}

As was mentioned, in order to allow that the \dS/\AdS\ spacetimes may
have different radii and the realistic cosmological constant may
contain  a variable part in addition to the fundamental  one, some variable numbers may be introduced together with
the universal constant $l$.  On the other hand,  as long as domain $r_\pm(x)$ (\ref{eq:Rpm}) is
replaced by domain $\sigma_\pm(x)$ (\ref{eq:domains})  the
degenerate Einstein manifold $M_\pm$ of $\Lambda_\pm=\pm 3 l^{-2}$
becomes  the Beltrami-\dS/\AdS\ space{time}
 with the same cosmological constant  and vice versa, respectively.
These  indicate that all these spacetimes and relevant
Poincar\'e, dual Poincar\'e, \dS\ and \AdS\ groups 
 can be dealt with simultaneously in the same
manner. In fact, these can all be done by introducing such a domain
and its absolute with a pair of numbers  $\{r,s\}$ that for their
four regions shown in (\ref{eq:ts}), it gives rise to four corresponding
domains, their absolutes of the spacetimes and their invariant
groups, respectively.
 Such a parameterized domain with  absolute is called that of
the Poincar\'e-\dS\ flow ${\cal F}_{r,s}$. Then the flow is just the
parameterized relativistic quadruple ${\frak Q}_{r,s}$.

\subsection{Domain, Absolute and Transformations of Poincar\'e-$\boldsymbol{dS}$ Flow}

Let us introduce the  domain  
and  the absolute 
of the Poincar\'e-\dS\ flow as follows
\be\label{eq:domaints}%
&\frak{D}_{r,s}:\quad \sigma_{r,s}(x):=\sigma_{r,s}(x, x)=r- s
l^{-2}xJ
x^t > 0,&\\\label{eq: sigmats=0}%
&\frak{B}_{r,s}=\partial(\frak{D}_{r,s}):\quad  \sigma_{r,s}(x)=r-s l^{-2}xJ x^t = 0.& %
 \ee%
It is clear that first for the fixed values of the numbers $\{r,s\}$  %
\be\label{eq:tsf}%
\{r,s\}|_{fixed}= \{1,0\}, \{0,1\}, \{1, 1\}, \{1,-1\}, \ee
 absolute
(\ref{eq: sigmats=0}) and its domain (\ref{eq:domaints})
 become that of Poincar\'e, dual  Poincar\'e and \dS/\AdS\ case,
 respectively. If one of the pair of numbers, \textit{i.e.}  $\{s\}$, is allowed to
 change its value in four regions 
 as follows
 \be\label{eq:ts}%
\{r,s\}\in \{\{1\},\{0\}\}, \{\{0\},(0,+\infty)\},
\{ \{1\},(0,+\infty)\}, \{\{1\},(-\infty,0)\}, %
\ee%
 the four types of  domains and absolutes can also be reached for  invariant group
${\cal P}/{\cal P}_2/{\cal D}_+/{\cal D}_-$, respectively.
  It is clear that
not only (\ref{eq: sigmats=0}) contains the absolutes of ${\cal P},
{\cal P}_2, {\cal D}_\pm$ for four pairs of fixed value of $\{r,s\}$
in (\ref{eq:tsf}), but also four types of the absolutes of ${\cal
P}, {\cal P}_2, {\cal D}_\pm$ with four different
regions (\ref{eq:ts}) of $\{r,s\}$, respectively. Namely,
\be\label{eq:4regions}%
\sigma_{r,s}(x)=r-sl^{-2}xJx^t=0: \ \ 
\left\{\begin{split}%
&  \xi^4=0, \ \ \quad r=1,\ \ s=0,\\
& r_\pm(x)=0, \ \ r= 0, \ \ s > 0,\\
& \sigma_\pm(x)=0,\ \ r=1,\ \  s\gtrless 0. \end{split}\right.
\ee%

In order to preserve 
absolute (\ref{eq: sigmats=0}) or (\ref{eq:4regions}), 
\LFT s (\ref{eq:LFT}) must reduce to their subset with the
following conditions
\begin{equation}
s A J A^t - r c^tc  = s J, ~ s A J b^t - r c^td = 0, ~ r d^2 - s b J
b^t = r. \label{eq:transfts}
\end{equation}
 Setting $b = - r
l^{-1} a A$ with respect to  the point $A(a^\mu)\in\frak{D}_{r,s}$
in  domain (\ref{eq:domaints}) of the flow ${\cal F}_{r,s}$, it
follows
\begin{equation}\label{eq:transfts1}
s A J A^t - r c^tc  = s J, ~s r l^{-1} A J A^t a^t + r c^t d = 0, ~
r d^2 - s r^2 l^{-2} a A J A^t a^t  = r,
\end{equation}%
In terms of the standard method of  projective geometry, especially
Hua's matrix analysis (see, \textit{e.g.} \cite{Hua, HuaC,Hua63,
HuaWan}), Eq. (\ref{eq:transfts1}) can be solved to get the
transitive form of the parameterized \LFT s of the flow
\begin{equation}\label{eq:LFTts}
\begin{split}
T_{r,s}:\quad x \rightarrow \tilde{x}
&= ( 1 - s r l^{-2} a J a^t )^{\frac 1 2} \d {x - r a }{1 - s l^{-2} x J a^t } A \\
A &= L + s r l^{-2}[(1-s r l^{-2} a J a^t)^{\frac 1 2} + (1 - s r
l^{-2} a J a^t)]^{-1} J a^t a L,
\end{split}
\end{equation}
and all these parameterized \LFT s form an invariant group.
{For four regions of $\{r, s\}$ in Eqs. (\ref{eq:tsf}) and (\ref{eq:ts}), \omits{\textit{i.e.}%
\be\nno
&\{r,s\}|_{fixed}= \{1,0^\pm\}, \{0^+,1\}, \{1, 1\}, \{1,-1\},
&\\\nno
 &\{r,s\}\in \{(0,+\infty),\{0^\pm\}\}, \{\{0^+\},(0,+\infty)\},
\{ (0,+\infty),(0,+\infty)\}, \{(0,+\infty),(-\infty,0)\},& %
\ee}%
it follows Poincar\'e
 transformations (\ref{eq:ISO(1,3)}) as  $s=0$, dual  Poincar\'e
 transformations in (\ref{eq:LFTP2})
 as $r = 0$ with length parameter $l |s|^{-1/2}$,
  and \dS/\AdS\ transformations  (\ref{eq:SO14,23}) as $s\gtrless 0$
  of radius  $l |s|^{-1/2}$, respectively.}
The infinitesimal generators of \LFT s (\ref{eq:LFTts}) read
\begin{equation}
\begin{split}
{\H}^{r,s} = r \r_t - s \nu^{2} t x^\alpha \r_\alpha,&\quad
{\P}^{r,s}_{\ i} = r \r_i + s l^{-2} x^i x^\alpha \r_\alpha,\\
{\K}^{r,s}_{\ i} 
={\bf K}_i,&\quad
\J^{r,s}_i=\J_i. 
\end{split}
\end{equation}
They satisfy
\begin{equation}
\begin{split}
[ {\bf P}^{r,s}_{\ i}, {\bf P}^{r,s}_{\ j}]=-rs l^{-2} \epsilon_{ij}^{\, \, \, k}{\J}_{k}, \quad [{\rm H}^{r,s}, &{\bf P}^{r,s}_{\ i} ] = r s \nu^2 {\bf K}_i, \quad [ {\bf P}^{r,s}_{\ i}, {\bf K}_j ] = \delta_{ij} c^{-2} {\rm H}^{r,s}, \\
[{\rm H}^{r,s}, {\bf K}_i ] = {\bf P}^{r,s}_{\ i},&\quad [ {\bf K}_i, {\bf K}_j  ] = r s c^{-2} \epsilon_{ij}^{\ \,k}{\J}_{k},\\
[\J_i, {\H}^{r,s}]=0, \quad [ {\J}_{i}, &{\bf P}^{r,s}_j ]={ -}\epsilon_{ij}^{\ \,k}{\P}^{r,s}_{k},\quad [\J_i, \K_j]={ -}\epsilon_{ij}^{\ \,k}{\K}_{k}. 
\end{split}
\end{equation}
For four regions of $\{t, s\}$ in Eqs.
(\ref{eq:tsf}) and (\ref{eq:ts}), it follows Poincar\'e,
 dual  Poincar\'e 
  and \dS/\AdS\ algebra, 
  respectively.

\subsection{Spacetime, Inertial Motion and Simultaneity of  Poincar\'e-$\boldsymbol{dS}$ Flow}

\omits{now find  the metric invariant under \LFT s (\ref{eq:LFTts})
of the Poincar\'e-\dS\ flow. } 
Let us consider a  line like (\ref{eq:L}) connecting two points
$(x_1,x_2)$ in
domain (\ref{eq:domaints}) %
with $x=x_1, x_2$ for $\tau = 0, 1$, respectively. The line
intersects absolute (\ref{eq: sigmats=0}) at two points $(\tau_+,
\tau_-)$
\begin{equation}
\tau_+ + \tau_- = - \d{2x_1 J (x_2 -
x_1)^t}{(x_2-x_1)J(x_2-x_1)^t}\quad \tau_+\tau_-= \d {s x_1 J
x_1^t - r l^2}{s(x_2-x_1)J(x_2-x_1)^t}.
\end{equation}
Thus, for four events with $\tau=(\tau_+, 1, \tau_-, 0)$, a cross
ratio can be given. From the power-2 cross ratio invariant, it
follows the interval between a pair of events $A(a^\mu)$ and
$X(x^\mu)$
  and the
lightcone with the vertex at $A(a^\mu)$ as  follows,
respectively
 \be\label{eq:Dltts}%
 &{\Delta}_{r,s}^2(A,
X) =s
l^{-2}\{\sigma_{r,s}^{-1}(a)\sigma_{r,s}^{-1}(x)\sigma_{r,s}^2(a,x)-1\},
&\\\label{eq:Blightconets} %
&{\cal C}_{rs}: \quad
\sigma_{r,s}^2(a,x) -\sigma_{r,s}(a)\sigma_{r,s}(x)=0.&%
\ee %
For 
two close events $X(x^\mu), X+dX(x^\mu+dx^\mu)\in \frak D_{r,s}$
(\ref{eq:domaints}),  the Beltrami metric of the flow follows from
(\ref{eq:Dltts})
\begin{equation}
\begin{split}
&ds^2_{r,s} = \sigma^{-1}_{r,s}(x) dxJ dx^t + s l^{-2}
\sigma^{-2}_{r,s}(x) (xJ dx^t)^2,\\
&{\bf g}_{r,s}:=({ g}_{r,s})_{\mu\nu}=
{\sigma^{-1}_{r,s}(x)} ( J \pm l^{-2} \d {J x^t x J }{\sigma_{r,s}(x)}).
\end{split} \label{eq:metricts}
\end{equation}
The contra-variant metric is
\begin{equation}\label{eq:contra-metricts}
{\bf g}^{-1}_{r,s} ={\sigma_{r,s}(x)} ( J \pm l^{-2} \d {J x^t x J }{\sigma_{r,s}(x)})^{-1}=
 \sigma_{r,s}(x)(J - \d s r l^{-2} x^t x).
\end{equation}
The last step of above equation can be obtained only when $r\neq 0$.
The metric (\ref{eq:metricts}) is degeneratedd for the region of
dual Poincar\'e case in (\ref{eq:4regions}), \textit{i.e.} $r=0,
s>0$. The corresponding contra-variant metric
(\ref{eq:contra-metricts}) is divergent and can be given formally
with the help of the flow when $r=0$. In terms of projective
geometry, $\d 1 r=\infty~(r=0)$ is well-defined. It is
straightforward  to get the Christoffel connection, Riemann
curvature, Ricci
 curvature
  and scale curvature of the flow, respectively, as follows 
 \be\nno
&\Gamma^{\ \ \lambda}_{r,s\ \mu\nu} =s l^{-2} \sigma_{r,s}^{-1} (
\delta^\lambda_\mu x_\nu + \delta^\lambda_\nu
x_\mu),&\\\label{eq:riccits}%
&R^{\,\,\,\mu}_{r,s \nu \la \si}(x)=s l^{-2}(g_{r,s
\nu\la}\dl^\mu_\si-g_{r,s \nu\si}\dl^\mu_\la)&\\\nno
&R_{r,s \mu\nu}(x) = {3s}l^{-2} g_{r,s \mu\nu}, \,~~\,R_{r,s}(x) =  {12 s}l^{-2}.&%
\ee %
Thus, the spacetime of the flow 
is an
Einstein manifold  with a parameterized cosmological constant
$\Lambda_s= {3s}l^{-2}$.

 It is also straightforward to check that the geodesic motion of the
 metric  (\ref{eq:metricts}) is indeed  inertial motion
 (\ref{eq:uvm}) of Newton's first law. In fact, the geodesic of
 the flow leads to the conservation law of the  (pseudo) 4-momentum of a particle
 with mass $m_{r,s}$ in the flow%
 \be\label{eq:geodts}%
 \frac{d}{ds_{r,s}}p^{\mu}_{r,s}=0,\quad
 p^\mu_{r,s}=m_{r,s}\sigma^{-1}_{r,s}(x)\frac{dx^\mu}{ds_{r,s}}.%
 \ee%
Then, it follows inertial motion (\ref{eq:uvm}) %
\be%
p^\mu_{r,s}=consts,\quad \frac{p^i_{r,s}}{p^0_{r,s}}=\frac{dx^i}{c
dt}=consts.%
\ee%
It is easy to show that the corresponding 4d angular momentum is
also preserved for the motion%
\be\label{eq:L4d}%
L_{r,s}^{\mu\nu}=x^\mu p^\nu_{r,s}-x^\nu p^\mu_{r,s}.%
\ee%
And there is generalized Einstein's formula for the particle%
\be\label{eq:EPLm}%
rE^2-r\vect{p}_{r,s}^2c^2- \frac{sc^2} {l^2} { \jmath_{r,s}}^2 +
\frac{sc^4}{l^2}{
k_{r,s}}^2=m_{r,s}^2 c^4. \ee%
{with energy $E=p_{r,s}^{\phantom{r,s}0}$, momentum
$p_{r,s}^{\phantom{r,s}i}$, $p_{r,s\:i}=\delta_{i
j}p_{r,s}^{\phantom{r,s}j}$, `boost' $k_{r,s}^{\phantom{r,s}i}$,
$k_{r,s\:i}=\delta_{ij}k_{r,s}^{\phantom{r,s}j}$ and 3-angular
momentum $\jmath_{r,s}^{\phantom{r,s}i}$,
$\jmath_{r,s\:i}=\delta_{ij} \jmath_{r,s}^{\phantom{r,s}j}$.}
 Similarly, it is also
true for the light signals 
in the flow.

 In order to make sense for inertial motion  (\ref{eq:uvm}) in
 measurements, the coordinate simultaneity should be defined as two events
 $A(a^\mu)$
 and $B(b^\nu)$ are simultaneous if and only if%
 \be\label{eq:a=b}
 a^0:=x^0(A) =x^0(B)=:b^0.%
 \ee%
This simultaneity leads to  the following $1+3$ split of metric
(\ref{eq:metricts})
\begin{equation}\label{eq:3+1}%
ds^2_{r,s} = N^2_{r,s} (dx^0)^2 - h_{r,sij}(dx^i + N_{r,s}^i
dx^0)(dx^j + N_{r,s}^j dx^0).
\end{equation}
Here the lapse function $N_{r,s}$, shift vector $N^i_{r,s}$, and
induced 3-geometry $h_{r,s ij}$  on 3-hypersurface $\Sigma_{r,s c}$
 read \be\nno
&h_{r,s ij} = -g_{r,s ij} = \sigma_{r,s}^{-1}(x)\delta_{ij}- s
\sigma_{r,s}^{-2}(x) l^{-2} x_i x_j& \\
&- h_{r,s ij} N_{r,s}^j = g_{r,s 0i} = s \sigma_{r,s}^{-2}(x)
l^{-2}x_0 x_i ,&\\\nno%
 &N^2_{r,s} - h_{r,s ij}N_{r,s}^iN_{r,s}^j
= g_{r,s 00} = \sigma_{r,s}^{-1}(x) + \sigma_{r,s}^{-2}(x) l^{-2}
x_0 x_0.&
\ee%
 The inverse of $h_{r,s ij}$ reads in {the $1+3$ split of
 contra-variant metric}
\begin{equation}
h_{r,s}^{ij} = \sigma_{r,s}(x)(\delta^{ij} + s l^{-2} x^i x^j (r - s
l^{-2} (x^0)^2)^{-1}).
\end{equation}
And the shift vector ${N}^i_{r,s}$ and the lapse function $N_{r,s}$
reads, respectively
\be%
{N}^i_{r,s}=\d { s x^0 x^i }{ r l^2 - s (x^0)^2}, \quad%
N_{r,s}^2 = \d {r l^2}{\sigma_{r,s}(x)(rl^2 - s(x^0)^2)}. \ee%

It is true that for four regions (\ref{eq:tsf}) and (\ref{eq:ts}) of
the numbers  $\{r,s\}$, the flow reaches that of Poincar\'e, dual
Poincar\'e, \dS\ and \AdS\ with different radius, respectively.

In particular, for the dual  Poincar\'e case,  degenerated metric
(\ref{eq:gMpm}) and its formal inverse (\ref{eq:hMpm}) can also be
defined as 
of 
the numbers  $r =0, s> 0$. And from $1+3$ split (\ref{eq:3+1}),
it follows that the non-degenerate part of degenerated metric
(\ref{eq:gMpm}) is a Beltrami metric of a 3d sphere/hyperboloid of
radius $l |s|^{-1/2}$ for space/time-like domain $\dot R_\pm$ at the origin
with the compact one $\bar C_O=\partial \dot
R_\pm$ in Eq. (\ref{eq:CObar}), and the non-degenerate part of
formal inverse (\ref{eq:hMpm}) is a time axis, respectively.

\subsection{Embedding Picture of  Poincar\'e-$\boldsymbol{dS}$ Flow}

Since the above 4d spacetime $S^4_{r,s}$ of the flow is constant curvature  formally, it can be embedded in a parameterized 5d flat
spacetime $M^5_{r,s}$ for $r\neq 0,s\gtrless 0$ of \dS/\AdS\ case,
respectively, as follows
\begin{equation}\label{eq:Sts}
S^4_{r,s}:\quad { |s|} \eta_{\mu\nu} \xi^\mu \xi^\nu - { \d {|s|}
s}r  (\xi^4)^2 = -\d {l^2} { s},
\end{equation}
which can be written as %
\be\label{eq:xiAts}%
 |s|\eta_{rs AB}\xi^A\xi^B=-\d {l^2} {s}, \quad {\cal J}_{r,s}:=\eta_{rs AB}={diag(J, -\d {r} s )}.%
\ee%
On $S^4_{r,s}\subset M^5_{r,s}$, there is  a parameterized metric
\begin{equation}\label{eq:dsSts}%
ds_{5}^2 = {|s|} \eta_{\mu\nu} d\xi^\mu d\xi^\nu - {\d {r}{s}|s|} (d
\xi^4)^2={ |s|}\eta_{rs AB}d\xi^A d\xi^B,
\end{equation}
which satisfies
\begin{equation}
ds^2_5 =ds^2_{r,s}.
\end{equation}%
It is clear that they are invariant under a subgroup ${\cal S}_{rs}$
of \LFT s
(\ref{eq:LFT}) consists of all $T_{r,s}$ satisfying %
\be\label{eq:Tts}%
T_{r,s}{\cal J}_{r,s}T_{r,s}={\cal J}_{r,s},\quad \forall
T_{r,s}={\cal
S}_{rs}.%
\ee%
These conditions are equivalent  to Eqs. (\ref{eq:transfts}). In
terms of the Beltrami coordinates like (\ref{eq:BLxU4}), Eqs.
(\ref{eq:Sts}) or (\ref{eq:xiAts}) and (\ref{eq:dsSts}) become
domain condition (\ref{eq:domaints}) and Beltrami metric
(\ref{eq:metricts}) of the flow, respectively. And transformations
$T_{r,s}$ in (\ref{eq:Tts}) are just that of \LFT s in
(\ref{eq:LFTts}). It should be noted that all matrixes $T_{r,s}$
satisfying (\ref{eq:Tts}) do form a Lie group with two numbers
$\{r,s\}$ (see, \textit{e.g.} \cite{ELflow}).

 We can also consider a  kind of uniform  great `circular' motions on (\ref{eq:Sts})
  for a
free massive particle with mass $m_{r,s}$ defined by  a conserved
5d  angular momentum, \textit{i.e.} %
\be\label{eq:dLts=0}%
&\frac{d{\cal L}_{r,s}^{AB}}{ds_{5}}=0,&\\\label{eq:Lts}%
& {\cal L}_{r,s}^{AB}=m_{r,s}|s|(\xi^A\frac{d\xi^B}{ds_{5}}-
\xi^B\frac{d\xi^A}{ds_{5}}).&
 \ee%
 And there is an Einstein-like formula for the particle:
 \be\label{eq:L^2}%
-\frac{s}{2 l^2}\eta_{rs AC} \eta_{rs BD}{\cal L}_{r,s}^{AB}{\cal L}_{r,s}^{CD}=m_{r,s}^2.%
\ee%
It is 
invariant under linear transformations $T_{rs}$ (\ref{eq:Tts}) of
${\cal S}_{rs}$ group of the flow.

 In terms of the Beltrami coordinates like (\ref{eq:BLxU4}), the parameterized Beltrami metric (\ref{eq:metricts}) follows from (\ref{eq:dsSts}) and
the uniform `great circular' motion (\ref{eq:dLts=0}) turns to the
geodesic motion in Eq. (\ref{eq:geodts}) and the conserved 5d
angular momentum in (\ref{eq:Lts}) becomes the conserved 4d
pseudo-momentum in Eq.  (\ref{eq:geodts}) and 4d angular momentum%
\be\label{eq:ptsLts}%
p^\mu_{r,s}=m_{r,s}\sigma^{-1}_{r,s}(x)\frac{dx^\mu}{ds_{r,s}}={ \d
{1} l}{\cal
L}_{r,s}^{4\mu},\quad L_{r,s}^{\mu\nu}={\cal L}^{\mu\nu}_{r,s}.%
\ee%
And Einstein-like formula (\ref{eq:L^2}) becomes the generalized
Einstein's formula (\ref{eq:EPLm}). %

In order to make sense for the uniform great circular motion
(\ref{eq:dLts=0}), simultaneity should be defined on the space{time}
${S}^4_{r,s}$ (\ref{eq:Sts}) of the flow.  For a pair of
simultaneous events $(P(\xi_P), Q(\xi_Q))$, their time-like
coordinates $\xi^0$, which is corresponding to the common
inhomogeneous time coordinate
$x^0$, should be equal to each other. Namely,%
 \be\label{eq:simtH}%
 \xi_P^0=\xi_Q^0.%
 \ee%
This simultaneity is the same with respect to the proper-time
simultaneity on $S^4_{r,s}\subset M^5_{r,s}$.
Then a simultaneous 3-space of $\xi^0=const$ reads: 
\be\label{eq:s3}
&{ |s|}\delta_{r,sIJ}\xi^I\xi^J=l^2{s^{-1}}+{ |s|}(\xi^0)^2,~~ I, J=1,\cdots, 4;&\\\nonumber
&dl^2|_{\xi^0=const}={ |s|}\delta_{r,sIJ}d\xi^I d\xi^J, ~~\delta_{r,sIJ}=diag({ I^3, \d {r} s}).& %
\ee %
For a kind of ``observers" ${\cal O}_{S^4_{r,s}}$ at the spacial
origin $O|\xi^\alpha=0$, where $\alpha$ takes three among
$1,\cdots,4$, it is a 3d hypersurface of variable radius $(l^2/{
s}+{ |s|}(\xi^0)^2)^{-1/2}$.

Along with $|\xi^0|$ enlarging, for the case of $r=1, s>0$,
\textit{i.e.} the \dS\ case, it is an expanding 3-sphere.  For the
case of $r=1,s<0$, \textit{i.e.}  the \AdS\ case, it is a
3-hyperboloid. Other two cases of Poincar\'e and dual  Poincar\'e in
Eqs. (\ref{eq:tsf}) and (\ref{eq:ts}) correspond to $r=1, s= 0$ and
$r = 0$, respectively.

\section{The Cosmological Meaning of Poincar\'e-$\boldsymbol{dS}$
Flow}\label{sec:Cosm}
 Let us now consider the cosmological meaning of the
 Poincar\'e-\dS\ flow. As was mentioned, As in the case of the
 Beltrami-\dS/\AdS\ spacetimes , there are two kinds of simultaneity, \textit{i.e.} the coordinate
 simultaneity and the propertime simultaneity, in the
Poincar\'e-\dS\ flow. Then it can also be studied the
Robertson-Walker-like cosmos of the flow, which coincides the
cosmological principle.

\subsection{Robertson-Walker-like Metric of Poincar\'e-$\boldsymbol{dS}$ Flow}

For the Robertson-Walker-like metric of the Poincar\'e-\dS\ flow, we
chose the simplest and most natural foliation. Namely, from the
parameterized Beltrami metric (\ref{eq:metricts}) of the flow to
this cosmic metric is just by changing the parameterized Beltrami
coordinate time to the proper-time as a co-moving or the cosmic time
and vise versa. Namely, As in the Beltrami model of \dS\
spacetime~\cite{BdS}, for the flow we can also introduce the
propertime $\tau$ as cosmic time that is related to Beltrami-time
coordinate $x^0$ by
\be\label{eq:tau}%
 x^0\sigma_{rs}^{-1/2}(x)=(\xi^0=) \d l {\sqrt{s|s|}} \sinh\left(\sqrt{ s }\tau/l\right).
\ee%
The metric (\ref{eq:metricts}) of the flow becomes
\begin{equation}\label{eq:RWmts}%
\begin{split}
&\qquad\quad ds^2_{r,s} = d \tau^2 -  \cosh^2\left(\sqrt{s}\tau/l\right)dl^2_0,\\%
& dl^2_0= \sigma_0^{-1}(x) \delta_{ij} dx^i dx^j - s
l^{-2}\sigma_0^{-2}(x)\delta_{ik}\delta_{jl}x^k x^l dx^i dx^j,\\
&\qquad\qquad\sigma_0(x)= r+sl^{-2}\delta_{ij}x^ix^j.
\end{split}%
 \end{equation}%
This is just a Robertson-Walker-like metric with the cosmic time.
The Beltrami metric (\ref{eq:metricts}) of the flow can also be
obtained from the Robertson-Walker-like metric (\ref{eq:RWmts}) by
replacing $\tau$ with Beltrami time via Eq. (\ref{eq:tau}).

It is clear that here the space foliation is a 3d sphere and  other flat or open space foliations may also be taken for other physical requirements.
\omits{In fact, the propertime $\tau$ is related to the Beltrami-time coordinate $x^0$ by %
so the Robertson-Walker-like metric (\ref{eq:RWmts}) can be gotten
directly by changing the Beltrami-time to the proper-time from the
Beltrami-metric (\ref{eq:metricts}) and vise versa.}

The
metric of 3d simultaneous hyper-surface with respect to  the { same} cosmic
time 
reads
\begin{eqnarray}\label{dSmosmos}%
dl_{\tau=const}^2 = R(\sqrt{s}\tau/l)dl_0^2=\cosh^2\left(\sqrt{s}\tau/l\right)dl^2_0.
\end{eqnarray}
It is clear that for $s>0$, metric (\ref{eq:RWmts})  describes a
Robertson-Walker-like \dS-space{time} as an accelerated expanding 3d
spherical cosmos with $dl^2_0$ as metric
 of a 3d sphere with radius $ls^{-1/2}$.
{ For the models with other space foliations, there are corresponding formulae, but the scalar factor $R(\sqrt{s}\tau/l)$ is the same.}

\subsection{Kinematics for the Universe  at Cosmic Scale}

From the above analysis, it follows that the Poincar\'e-\dS\ flow
links the \PoRcl\ and the cosmological principle by changing the
simultaneity. As was emphasized by Bondi~\cite{Bondi},
Bergmann~\cite{Bergmann} and Rosen~\cite{Rosen} long ago and implied
by Coleman,  Grashow~\cite{CG} and others recently, there is 
the puzzle between the \PoR\ and the cosmological principle in
Einstein's theory of relativity. However, due to the  extreme asymptotical behaviors of our universe, the flow is out of the puzzle.

 More importantly,
for the numbers  $\{r,s\}\in \{\{1\},(0,+\infty)\}$, 
the flow turns to the \dS\ case with variable
curvature radius $ls^{-1/2}$.  In fact, for the
Robertson-Walker-like \dS\ cosmos with a variable radius $l
s^{-1/2}$ for different $s$, there is a horizon with Hawking
temperature
 and non-gravitational entropy due to non-inertial effect~\cite{TdS}
 \be\label{entropy}%
S=\pi l^2s^{-1}
{c^3k_B G\hbar}=k_B\pi l^2s^{-1}\ell_P^{-2},
\ee%
where $k_B$ is  the
Boltzmann constant and $\ell_P$ the Planck length.
Obviously, for different value of $s$, it  works at cosmic scale with an accelerated
expanding spherical cosmos of radius $R$ if $l s^{-1/2}=R\simeq
\sqrt{3/\Lambda}$.

If $\Lambda$ is directly taken from observation without slightly variation, it gives rise
to an accelerated expanding
$S^3$ model of radius $R$ and entropy $S_R$ %
\be\label{S_R}%
R\simeq (3/\Lambda)^{1/2}\sim 13.7 Gly, ~ S_R=k_B\pi g^{-2},\, g^2:=(\ell_P/R)^2\simeq 10^{-122},%
\ee%
where $S_R$ may provide an upper entropy  bound for our universe. {
It is also important that the entropy bound is independent of the
space foliation~\cite{TdS}.} Here an important dimensionless
constant $g$ appears. It should characterize the dimensionless gauge
coupling in the gauge theory model of \dS\
gravity~\cite{PoI,Lu80,dual07,Vacuum,torsion}.

Since the time-arrow of the universe should coincide with the arrow
of cosmic time  axis of the  Robertson-Walker-like \dS\ cosmos,
which is  a  counterpart of the Beltrami-\dS\ space{time} with
Beltrami time axis related to the cosmic time axis by changing the
Beltrami-time to the proper-time as the cosmic time, the evolution
of our universe definitely indicate the existence of the
Beltrami-frame of inertia\cite{PoI,Lu80}. This manner for
determination of the Beltrami-frame of inertia is completely
different from that described by Einstein  and is  also away from
Einstein's ``an argument in a circle" for the \PoR~\cite{1922}.

In addition, with different $s$ the Robertson-Walker-like \dS\ cosmos may also provide an inflationary  phase near the
Planck length $\ell_P$ if %
\be\label{S_P}%
ls^{-1/2}\simeq\ell_P, \quad S_{\ell_P}=k_B\pi,%
 \ee%
or at other scales as the GUT scale for the inflation model. For the Planck scale, it may provide kinematically  a lowest entropy bound for the universe. Of course, other inflationary phase with different scale may also work with corresponding entropy bound. But, (\ref{S_P}) is the lowest one.  The relation between this inflationary  phase and the inflation model should be studied further.

What is real kinematics for the universe? Since the cosmological constant $\Lambda$ is positive, it must be the \dS\ \SR.

In fact, our theoretical analysis should also conclude that this is the case even if the cosmological constant could be unknown.
  It is important that our universe is always with entropy that is increasing and roughly described symmetrically by the cosmological principle. In addition, its kinematics should be based on the \PoR\ as well. Therefore, in the sense of the \PoRcl,
the cosmological principle and the principle of increasing entropy, the \dS\ \SR\  with double $[{\cal D}_+,{\cal
P}_2]_{D_+/M_+}$ should definitely provide new kinematics for our universe. While two other kinds of \SR\ of Poincar\'e/\AdS\ invariance cannot.
Thus, the conventional relativistic physics works %
 up to the bound (\ref{10-35})
 locally
  except for the both cosmic scale and near Planck scale.

In conclusion, the \dS\ \SR\ should at least describe physics kinematically   at the cosmic scale with the  entropy upper bound  (\ref{S_R}) for the accelerated expanding
universe. It
and may also work  for the beginning of our universe near  the Planck scale with the entropy lowest bound (\ref{S_P})  of
an inflationary  phase or at other scale for an inflation.

Since the Poincar\'e-\dS\ flow may  allow the \dS\ space{time} with different radius $l s^{-1/2}$, it may also allow correspondingly to divide the observed cosmological constant $\Lambda$ into a slightly variable part $\Lambda_v$ and the fundamental one $\Lambda_0$.  The latter  should be related to the value of the universal invariant constant $l$. Then one may simply take that $\Lambda$ is $\Lambda_s$ and $\Lambda_0$ relates to $l$ only, so  the (slightly) variable part reads%
 \be\label{Lambdav}%
\Lambda_v=\Lambda_s-\Lambda_0= {3(s-1)}l^{-2},\quad s\sim 1.
 \ee%
 In order to determine the concrete values of two parts or equivalently the value of $l$, one should refer to the detailed analysis on the observation data at least. 

\section{Concluding  Remarks}
The accelerated expanding universe strongly indicates that there are no longer Euclidean rigid ruler and ideal clock for physics at the cosmic scale characterized by the cosmological constant. As long as the Euclidean assumption for space and time
is given up,  Newton's inertial motion is allowed may reach the projective
infinity. Then,  the inertial motion group
$IM(1,3)\sim PGL(5,R)$ follows  based on the \PoRcl\, and there are more candidates of
metric geometry as  space and time 
 for kinematics.

With common
 Lorentz isotropy, there are three kinds of \SR\ as a triple associated with the relativistic quadruple ${\frak Q}_{PoR}=[{\cal P}, {\cal P}_2, {\cal
D}_+,{\cal D}_-]_{M/M_\pm/D_\pm}$. The quadruple can be parameterized as the
Poincar\'e-\dS\ flow  ${\cal F}_{r,s}$. \omits{In the quadruple
${\frak Q}_{PoR}$, the dual Poincar\'e group ${\cal P}_2$-invariant
degenerate Einstein manifold $M_\pm$ of $\Lambda_\pm=\pm3l^{-2}$ is
for the common space/time-like region $R_\pm$  of the  lightcone
$\bar C_O=\partial \dot R_\pm$ at origin on Minkowski/\dS/\AdS\
space{time} $M/D_\pm$, respectively. Both are associated to the \SR\
triple of three kinds of Poincar\'e/\dS/\AdS\ group ${\cal P}/{\cal
D}_\pm$ invariant special relativity. }Thus, three kinds of \SR\ of
Poincar\'e/dS/\AdS-invariance associated with the dual Poincar\'e
invariant degeneratedd spacetimes can be reached by the
Poincar\'e-\dS\ flow   ${\cal F}_{r,s}$ with a pair of numbers
$\{r,s\}$ of different values. In fact, the radii of the \dS/\AdS\
spacetimes are allowed to be different as $l |s|^{-1/2}$ and
correspondingly the cosmological constant  can also be changed as
$\Lambda_s$ like in Eq. (\ref{Lambdav}).

For the flow, there are the parameterized Beltrami-time simultaneity
and the parameterized inertial motions for a free particle with a
parameterized Einstein's formula. For different values of $\{r,s\}$,
they give rise to the ones of the Poincar\'e/dual
Poincar\'e/\dS/\AdS-invariance on  $M/M_\pm/D_\pm$, respectively. On
the other hand, with respect to  the parameterized propertime
simultaneity of the flow,  the  Beltrami metric of the flow should
turns to the  Robertson-Walker-like one and vice versa.  Thus, for
the  extremely asymptotical behavior of the universe,  the flow gets
ride of the puzzle between the \PoR\ and the cosmological
principle~\cite{Bondi,Bergmann,Rosen,CG} and the evolution of our
universe does fix all kinds of the inertial coordinate frames in the
flow just like the case of the Beltrami-\dS\ inertial
frames~\cite{PoI}.

The dual Poincar\'e kinematics can be studied more clearly as a
member 
of the Poincar\'e-\dS\ flow. Although there is no an independent
meaningful  4d ${\cal P}_2$-invariant kinematics, rather on a pair
of degenerated Einstein manifolds $M_\pm$ of $\Lambda_\pm$, it
always appears  as an associated partner at the common origin on
Minkowski/\dS/\AdS\ space{time} $M/D_\pm$ for the space/time-like
domain $\dot R_\pm$  of the compact lightcone $\bar C_O$ generated
by the \LFT s of dual Poincar\'e group  ${\cal P}_2$  from the
space/time-like region $ R_\pm$ of the lightcone $C_O$ at the common
origin, respectively.

In the sense that for the whole universe it holds the \PoRcl,
the cosmological principle and the principle of increasing entropy, the \dS\ \SR\  with double $[{\cal D}_+,{\cal
P}_2]_{D_+/M_+}$ should  provide consistent kinematics for  the cosmic scale physics with the  entropy upper bound  (\ref{S_R}). In addition, it may also work for physics near  the Planck length with the entropy lowest bound (\ref{S_P})  of
an inflationary  phase or at other scale. It is worthy while to study its relation to the inflation model further.
Since the cosmological constant is very tiny, the conventional relativistic physics can still work %
 up to the bound (\ref{10-35})
 locally
  except for the cosmic scale.
  It is clear that in order to determine the concrete values of a slightly variable part and the fundamental one of the cosmological constant or equivalently the value of $l$, the detailed analysis on the observation data are needed  at least. %

From the viewpoint of the \PoRcl, dynamics should coincide
with kinematics. Actually, for  Newton's second law with 
an integral curve diffeomorphic to the projective straight line for
the inertial motion of Newton's first law, the symmetry of such kind
of Newton's second law is still the inertial motion group
$IM(1,3)\sim PGL(5,R)$.   And gravity should be based on the
local-globalization of kinematics. Namely, to localize  the
kinematic symmetry and its spacetime in patches first and to transit
these patches together globally to form a kind of $1+3$d manifolds,
and to describe gravity with some gauge-like field equation
characterized by the dimensionless coupling constant $g$ in Eq.
(\ref{S_R}) (see, \textit{e.g.}
\cite{PoI,Lu80,dual07,Vacuum,torsion}).

Although there are important results for the Poincar\'e-\dS\ flow
and as was mentioned, the roles of  the flow are still mainly  in
mathematics. In particular, as the number $s$ varies, it   may
realize the concrete procedures of different contractions of the
universal constant $l$, although the constant $l$ is always
invariant here. Thus, one may also introduce one more real number to
realize the concrete procedures of the different contractions of
$c$, \textit{i.e.} $c\to \infty$ and $c\to 0$. Then  the method
employed and basic considerations here can be applied to other
geometrical and non-relativistic kinematic symmetries.

Of course, more physical meaning  of the flow with these numbers  and above  issues should be explored further.




\begin{acknowledgments}
\omits{This work is partly completed in ``Connecting Fundamental
Physics with Observations" programm, KITPC, CAS.} We would like to
thank Z. Chang, Y.B. Dai, S. Gong,   C.-G. Huang, K. Li, Q.K. Lu, Z.C. Lu,
 X.A. Ren, Y. Tian, Z.X. Wan, S.K. Wang, K. Wu,
X.N. Wu, Z. Xu,  X. Zhang and B. Zhou
for  discussions and comments. The work is partly supported by NSFC
Grants No. 10701081, 10975167, 10875129.

\end{acknowledgments}

\end{document}